# A model for the contraction kinetics of cytoskeletal gel slabs

Matteo Ferraresso[+], Mohammad Shojaeifard[+], Albert Kong, Mattia Bacca[*]

Mechanical Engineering Department, Institute of Applied Mathematics
School of Biomedical Engineering
University of British Columbia, Vancouver BC V6T 1Z4, Canada

[+]Equally contributed to the paper
[*]Corresponding author: mbacca@mech.ubc.ca

**Abstract**

Cytoskeletal gels are engineered prototypes that mimic the contractile behavior of a cell in-vitro. They are composed of an active polymer matrix and a liquid solvent. Their contraction kinetics is governed by two dynamic phenomena: *mechanotransduction* (molecular motor activation), and solvent *diffusion*. In this paper, we solve the transient problem for the simple case of a thin gel slab in uniaxial contraction under two extreme conditions: motor-limited or slow motor activation (SM) regime, and diffusion-limited or fast motor activation (FM) regime. The former occurs when diffusion is much faster than mechanotransduction, while the latter occurs in the opposite case. We observe that in the SM regime, the contraction time scales as $t/t_0 \sim (\lambda/\lambda_0)^{-3}$, with $t_0$ the nominal contraction time, and $\lambda$ and $\lambda_0$ are the final and initial stretches of the slab. $t_0$ is proportional to $1/\dot{w}$, where $\dot{w}$ is the average mechanical power generated by the molecular motors per unit reference (dry polymer) volume. In the FM regime, the contraction time scales as $t/t_1 \sim (1-\lambda/\lambda_0)^2$, with $t_1$ the nominal contraction time, here proportional to the ratio $L^2/D$, where $L$ is the reference (dry polymer) thickness, and $D$ is the diffusivity of the solvent in the gel. The transition between the SM and FM regimes is defined by a characteristic power density $\dot{w}^*$, where $\dot{w} \ll \dot{w}^*$ gives the SM regime and $\dot{w} \gg \dot{w}^*$ gives the FM regime. Intuitively, $\dot{w}^*$ is proportional to $D/L^2$, where, at a given power density $\dot{w}$, a thinner gel slab (smaller $L$) or including smaller solvent molecules (higher $D$) is more likely to be in the SM regime given that solvent diffusion will occur faster than motor activation.

*Significance*

The proposed model describes the contraction kinetics of cytoskeletal gel slabs under the extreme conditions of very fast and very slow molecular motor activation. The former regime is limited by solvent diffusion, while the latter is limited by the reaction kinetics of motor activation (mechanotransduction). We provide simple time scalings for both regimes, based on leading order terms, and identify the transition power density (generated by motors) between these regimes.

**Introduction**

Cytoskeletal gels are in-vitro prototypes used to study the active mechanical behavior of the cytoskeleton, the structural backbone of a biological cell [1-4]. These are active polymer networks (actin) diluted inside an aqueous solvent (cytosol). The network is activated by molecular motors and transduces chemical energy into mechanical work, mimicking the contractile behavior of a biological cell, while excluding other cytoskeletal structures such as microtubules, organelles, the nucleus and lipid membranes [5-6]. In the contractile gel system, contractile motors, representing acto-myosin II complexes (non-muscle), generate active forces and stiffens the polymer network (actin) by either increasing the crosslink density between chains or by shortening the chains [7]. These two processes consume ATP (Adenosine Triphosphate) through hydrolysis powering the contractile motors, which transduce chemical energy to mechanical contractile work. Gel stiffening increments the osmotic pressure of the solvent and motivates its outflow, which finally accommodates macroscopic contraction [7-8].

The first models describing cytoskeletal gel swelling and contraction approached the problem from a hydrodynamic standpoint, where transient force dipoles, generated by molecular motors, induce contractile (active) stresses [9]. This approach is used to replicate the active behaviour of the cytoskeleton macroscopically, where molecular motors need to constantly consume power to maintain the system at the desired out-of-equilibrium state [10]. The main limitation of force-dipole models lies in the absence of microstructural dependence of active forces, which emerge from polymer network stiffening, and the assumption that actin filaments must be taught.

Recent studies have focused on the dynamics of network rearrangement and evolution, specifically its impact on the mechanical response. In coupling biochemical signals with reaction kinetics and actin chain evolution/activation, studies such as Deshpande et al. [11-12] and Liu et al. [13] developed continuum models that incorporate microscopic actin network dynamics. Such models first describe the reaction kinetics of actin chain polymerization, network assembly and contraction activation based on calcium signalling. A stress-strain rate relationship similar to Hill's stress-velocity curve is then implemented to connect phosphorylation levels to the mechanical stress response [14]. The final stress relationship is the combination of neo-Hookean hyper-elasticity and stress-strain rate contractile actin network dependencies. The main drawback to this approach stems from the inability to directly analyze the energetic landscape.

The most recent models use passive bi-phasic polymer gel platforms to study both the elastic and viscous responses. This physically more representative model is founded on thermodynamic properties of chains and solvent and, where the total free energy of the system is defined as a combination of chain strain energy and chain-solvent mixing energies [15-16]. The free energy expression is derived from microscopic parameters that all have a physical foundation, and therefore can all be quantified. A network-specific solvent diffusivity then introduces dissipation of energy as solvent flows out to achieve contraction [17]. Therefore, in this model both a solid (chains) and liquid (solvent) phase, interacting, mould the behaviour of the gel. Active polymer gels then evolved such models to study dynamic gel contraction and swelling [8, 18]. These gel models employ the backbone of passive bi-phasic gels but introduce evolution of the microstructure. The activation of contractile molecular motors such as myosin II increases the chain strain energy, leading to a stiffening of the chain network, inducing solvent outflow and gel contraction [19-20]. Other studies focus on the transient response of the gel during contraction, after stiffening due to chain density increase [8, 21] or local growth modeled as a thermal expansion [22-23]. They analyze different geometries and study solvent outflow in different configurations, while comparing these results with experiments.

The cytoskeletal mechanical response, and particularly contraction, is crucial to cell scale functions such as growth, division and development, but it also has implications in tissue wide responses. Understanding the mechanisms through which cells and tissues contract is of great interest; in fact, multiple studies focus on mathematical modelling to investigate such mechanisms. This paper builds on current cytoskeletal modelling approaches involving a polymeric gel representation of the system and studies two contractile regimes. Following [7], we model active gel stiffening as a combination of two effects, namely an increment in crosslink density, via dynamic crosslinking motors (DCM), and a reduction of the average chain length of the polymer backbone, via chain shortening motors (CSM). [7] discussed the energetic tradeoff for contraction analyzing the distinct action of both DCM and CSM mechanisms separately. The energetics of contraction was also analyzed in the two limit cases of very fast motor activation (FM), where the contraction kinetics are controlled by solvent diffusion, and very slow motor activation (SM), where the contraction kinetics are controlled by motor activation, or by the ATP hydrolysis rate. [7], however, only analyzed the energetic requirement for contraction for a given contraction amount but did not study the kinetics of contraction. In this work, we analyze the kinetic problem of contraction in the two limit cases of FM and SM as generated distinctively by DCM and CSM mechanisms.

Our model system observes the contraction of a simplified gel specimen geometry, namely a gel slab, where solvent flow, accommodating contraction, only occurs in one direction. We observe that in the SM regime, the contraction time $t$ scales as $t \sim t_0 (\lambda/\lambda_0)^{-3}$, with $\lambda/\lambda_0$ the ratio of final slab stretch (thickness) to the initial one, and $t_0$ the SM nominal contraction time proportional to $1/\dot{w}$, with $\dot{w}$ the motor power generation per unit reference volume (for reference state we adopt that of the dry polymer, prior to swelling due to solvent absorption). In the FM regime, we instead observe the scaling $t \sim t_1 (1 - \lambda/\lambda_0)^2$, where $t_1$ is the FM nominal contraction time, proportional to $L^2/D$, with $L$ the slab thickness in the reference (dry polymer) state and $D$ the diffusivity of the solvent. We ultimately identify the characteristic power generation $\dot{w}^*$ for which $\dot{w} \gg \dot{w}^*$ provide the FM regime, while $\dot{w} \ll \dot{w}^*$ gives the SM regime. Intuitively, such a characteristic power generation scales with the ratio $D/L^2$.

**Active Gel Model**

Let us define the reference state of the gel as that at which the polymer is dry, prior to swelling due to the presence of the solvent. The free energy density of the gel, per unit reference volume, is $\psi \, kT/\Omega$, with $k$ Boltzmann's constant, $T$ temperature, and $\Omega$ the average volume of a molecule of solvent. The dimensionless free energy density $\psi$ is given by the relation [7-8, 17-18]

$$\psi = \psi_e + \psi_m + p(1 + \Omega C - J) \tag{1}$$

where $\psi_e \, kT/\Omega$ is the elastic strain energy density of the polymer, $\psi_m \, kT/\Omega$ is the free energy density of mixing between polymer and solvent, $p \, kT/\Omega$ is the total pressure of the gel, $C$ is the solvent concentration in molecules of solvent per unit reference (polymer) volume, and $J = dV/dV_0$ the volumetric swelling of the gel, with $dV$ and $dV_0$ the unit volumes in the current and reference states, respectively. In Eq. (1), the dimensionless pressure $p$ enforces the incompressibility constraint

$$J = 1 + \Omega C \tag{2}$$

and thus, acts as a Lagrange multiplier. The dimensionless free energy density is

$$\psi_e = \frac{n}{2}[\lambda_M^{-2} \sum_{i=1}^{3} \lambda_i^2 - 2\ln(J\lambda_m^{-3}) - 3] \tag{3}$$

where $n = \Omega N$, with $N$ the crosslink density per unit reference volume, $\lambda_i$ are the principal stretches of the polymer, $J = \lambda_1 \lambda_2 \lambda_3$, and $\lambda_M$ is the micro contraction of the polymer [7]. Eq. (3) describes the configurational elasticity of a polymer network composed of $N$ Gaussian, slack, chains in the unit reference volume. The dimensionless mixing energy density is

$$\psi_m = \Omega C \ln\left(\frac{\Omega C}{1+\Omega C}\right) - \frac{\chi}{1+\Omega C} \tag{4}$$

where $\chi$ is the Flory parameter [15-16]. Eq. (4) emerges from the Flory-Huggins' theory, where the first term on the right-hand side is the configurational (entropic) energy of the solvent, and the second term describes the steric interactions between polymer chains (P) and solvent molecules (S). In this term, a positive $\chi$ describes P-S repulsion, while a negative $\chi$ describes P-S attraction.

In the next sections we explore two extreme cases, namely, (i) motor-limited contraction, in which the MM activation rate is much slower than the rate of solvent diffusion; and (ii) diffusion-limited contraction, in which the diffusion rate is much slower than MM activation.

As depicted in Figure 1, the model system is defined by unidirectional contraction of the gel slab along the $z$-axis. The initial conditions are defined by the swollen gel in chemo-mechanical equilibrium with the environment, where the principal stretch in all directions is $\lambda_0 = J_0^{1/3}$, with $J_0$ the initial swelling ratio. The principal stretch in direction $Z$ is

$$\lambda = \frac{\partial \tilde{z}}{\partial \tilde{Z}} \tag{5}$$

where $\tilde{z} = z/L$, and $\tilde{Z} = Z/L$. The principal stretch in the transverse directions is fixed as $\lambda_0$, so that

$$J = \lambda \lambda_0^2 \tag{6}$$

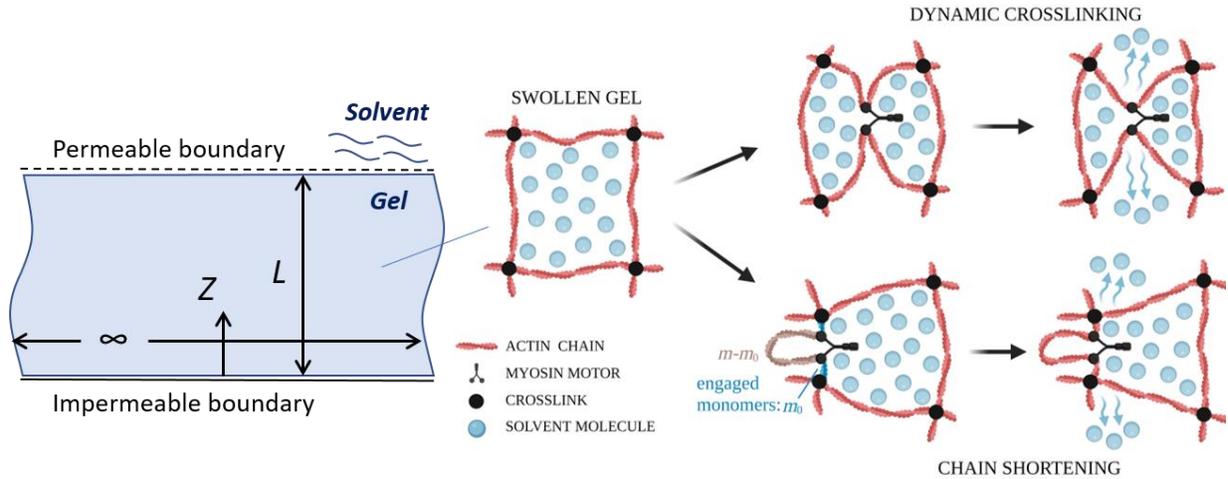

**Figure 1**: Schematics of the gel specimen in uniaxial contraction (*left*); the gel contracts via molecular motor activity, which can be described as dynamic crosslinking (increment in crosslink density) or chain shortening (reduction of engaged monomers between crosslinks) (*center-right*). The solvent diffuses out of the gel through the permeable boundary. The configuration and coordinate system sketched in this figure is that in the reference state (dry polymer), and defined by the coordinate $Z$, varying between $0$ and $L$.

*Mass conservation* inside the gel specimen imposes

$$\frac{\partial J}{\partial t} = -\frac{1}{L}\frac{\partial H}{\partial \tilde{Z}} \tag{7}$$

where $H$ is the solvent flux. The latter is defined by Fick's law [8, 17-18], which, using Eq. (2), writes as

$$H = -\frac{D}{L}\frac{J-1}{\lambda^2}\frac{\partial \mu}{\partial \tilde{Z}} \tag{8}$$

where $D$ is the diffusivity of the solvent inside the gel, and $\mu\, kT$ is the chemical potential of a solvent molecule in the gel, given by

$$\mu = \frac{1}{\Omega}\frac{\partial \psi}{\partial C} \tag{9}$$

By substituting Eq. (1), (2), and (4) into (9), we have

$$\mu = \frac{\chi+J}{J^2} + \ln\left(1 - \frac{1}{J}\right) + p \tag{10}$$

By substituting Eq. (10) into (9), and the result into (8) we can describe solvent diffusion, where the condition

$$\mu = \mu_e, \text{ at } \tilde{Z} = 1 \tag{11a}$$

applies at the permeable boundary, and

$$H = 0, \text{ at } \tilde{Z} = 0 \tag{11b}$$

applies at the impermeable boundary. Here, $\mu_e = p_e$ is the dimensionless chemical potential of the solvent bath where the gel specimen is immersed, and $p_e$ is its dimensionless hydrostatic pressure. To find the distribution of pressure $p$ in time, one has to solve the coupled mechanical problem.

The Cauchy (true) stress in the $i$-th principal direction is $\sigma_i\, kT/\Omega$, where

$$\sigma_i = \frac{\lambda_i}{J}\frac{\partial \psi}{\partial \lambda_i} \tag{12}$$

Substitution of Eq. (1) and (3) into (12) gives the dimensionless principal Cauchy stresses as

$$\sigma_i = \frac{n}{J}\left(\frac{\lambda_i^2}{\lambda_M^2} - 1\right) - p \tag{13}$$

*Equilibrium* at the permeable boundary imposes

$$\sigma_Z = -p_e, \text{ at } \tilde{Z} = 1 \tag{14}$$

while mechanical equilibrium inside the gel imposes

$$\frac{\partial \sigma_Z}{\partial \tilde{Z}} = 0 \tag{15}$$

Eq. (15), together with Eq. (14), imposes the uniform stress $\sigma_Z = -p_e$ in the gel. This, substituted into Eq. (13), gives

$$\Delta p = \frac{n}{J}\left(\frac{\lambda^2}{\lambda_M^2} - 1\right) \quad (16)$$

with $\Delta p = p - p_e$ the total pressure relative to that of the external solvent. From Eq. (16), (10) can be rewritten as

$$\Delta \mu = \frac{\chi + J}{J^2} + \ln\left(1 - \frac{1}{J}\right) + \frac{n}{J}\left(\frac{\lambda^2}{\lambda_M^2} - 1\right) \quad (17)$$

with $\Delta \mu = \mu - p_e$ the relative chemical potential. We can then substitute $\mu$ with $\Delta \mu$ in (8), together with Eq. (6), to obtain

$$H = -\frac{D}{L}\frac{\lambda\lambda_0^2-1}{\lambda^4\lambda_0^2}\left[\frac{\lambda\lambda_0^2+2\chi(1-\lambda\lambda_0^2)}{\lambda\lambda_0^2(\lambda\lambda_0^2-1)} + n\left(\frac{\lambda^2}{\lambda_M^2}+1\right)\right]\frac{\partial \lambda}{\partial \tilde{Z}} \quad (18)$$

where we have assumed that both microstructural parameters $n$ and $\lambda_m$ are uniform in the gel. This hypothesis is supported by the assumption of a uniform distribution of mechanical work, per unit reference volume, provided by the motors. Now, by substituting Eq. (6) and (18) into (7) we obtain the partial differential equation (PDE)

$$\frac{\partial \lambda}{\partial \tilde{t}} = \frac{\partial}{\partial \tilde{Z}}\left\{\frac{\lambda\lambda_0^2-1}{\lambda^4\lambda_0^4}\left[\frac{\lambda\lambda_0^2+2\chi(1-\lambda\lambda_0^2)}{\lambda\lambda_0^2(\lambda\lambda_0^2-1)} + n\left(\frac{\lambda^2}{\lambda_M^2}+1\right)\right]\frac{\partial \lambda}{\partial \tilde{Z}}\right\} \quad (19)$$

in the function $\lambda(\tilde{t},\tilde{Z})$ governing the contraction kinetics, where $\tilde{t} = t/t_d$ is the dimensionless time with $t_d = D/L^2$ the characteristic diffusion time. The initial conditions for this PDE are given by $\lambda(0,\tilde{Z}) = \lambda_0$, while its boundary conditions are given by Eq. (11). In the latter, Eq. (11a) provides a condition for $\lambda(\tilde{t},1)$, which simply gives $\Delta \mu = 0$ substituted from Eq. (17). Eq. (11b), on the other hand, provides simply the condition $\partial \lambda(\tilde{t},0)/\partial \tilde{Z} = 0$.

**Results and Discussion**

*Chemo-mechanical equilibrium*

The permeable surface of the gel is in chemo-mechanical equilibrium with the solvent buffer it is immersed through Eq. (11a) (chemical) and (14) (mechanical). The gel slab is in chemo-mechanical equilibrium if in the absence of solvent flow, *i.e.*, when $H = 0$. This, from Eq. (8), (11a), and (17), imposes $\Delta \mu = 0$, which, together with Eq. (6), gives the chemo-mechanical equilibrium condition

$$n = \frac{1 + \chi\,\lambda^{-1}\lambda_0^{-2} + \lambda\,\lambda_0^2 \ln(1 - \lambda^{-1}\lambda_0^{-2})}{1 - \lambda^2\lambda_M^{-2}} \quad (20a)$$

$$\sim \frac{1-2\chi}{2}\lambda_M^2 \lambda^{-3}\lambda_0^{-2} \quad (\lambda, \lambda_0 \gg 1) \quad (20b)$$

Eq. (20a) provides a finite $n$ only for $\lambda \geq \lambda_M$. In the passive configuration (0), prior to motor activation, the gel deformation is defined by $\lambda = \lambda_0$ and $J = J_0 = \lambda_0^3$. Also, the microstructure is defined by $n = n_0$, initial dimensionless crosslink density, and $\lambda_M = 1$. By substituting these quantities into Eq. (20) we obtain the initial chemo-mechanical equilibrium condition [7]

$$n_0 = \frac{1+\lambda_0^{-3}\chi+\lambda_0^3 \ln(1-\lambda_0^{-3})}{1-\lambda_0^2} \tag{21a}$$

$$\sim \frac{1-2\chi}{2}\lambda_0^{-5} \quad (\lambda_0 \gg 1) \tag{21b}$$

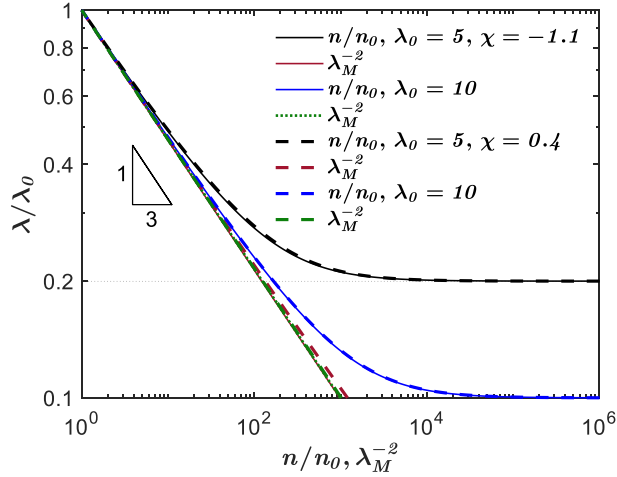

**Figure 2**: Chemo-mechanical equilibrium: Evolution of the stretch ratio $\lambda/\lambda_0$ as a function of crosslink density ratio $n/n_0$, for dynamic crosslinking motors (DCM), or as a function of $\lambda_M$, for chain shortening motors (CSM), for a Flory parameter $\chi = -1.1$ (polymer-solvent attraction) and $\chi = 0.4$ (repulsion), and for initial (swelling) stretch $\lambda_0 = 5$ and $10$.

The cytoskeletal gel prototype provided by [6] has $\lambda_0 = 10$, giving $J_0 = 1000$, and $n_0 = 1.61 \cdot 10^{-5}$, from which we can deduce the Flory parameter as $\chi = -1.1$, from Eq. (21). A negative $\chi$ indicate attractive interactions between polymer and solvent, as can be deduced from Eq. (4). Considering also the case of repulsive interactions, we adopt $\chi = 0.4$, from which, $\lambda_0 = 10$ and $J_0 = 1000$ give $n_0 = 1.01 \cdot 10^{-6}$ from Eq. (21). To also consider the effect of $\lambda_0$, we adopted $\lambda_0 = 5$, giving $J_0 = 125$, and $n_0 = 5.34 \cdot 10^{-4}$ ($n_0 = 3.42 \cdot 10^{-5}$) for $\chi = -1.1$ ($\chi = 0.4$).

Dynamic crosslinking motors (DCM) provide only an increment of the ratio $n/n_0$, while $\lambda_M$ remains unity. Chain shortening motors (CSM), on the other hand, evolve only $\lambda_M$, while $n/n_0$ remains unity. When only DCM are active, chemo-mechanical equilibrium defines the relation between $n/n_0$ and $\lambda/\lambda_0$ by equating Eq. (20) and (21), as plotted in Figure 2. When only CSM are active, chemo-mechanical equilibrium defines the relation between $\lambda_M$ and $\lambda/\lambda_0$ by equating Eq. (20) and (21), as plotted in Figure 2. In this figure, for both $n/n_0$ (DCM) and $\lambda_M^{-2}$ (CSM), we adopt $\chi = -1.1$ (attractive interactions between polymer (P) and solvent (S)) and $\chi = 0.4$ (repulsive P-S interactions), as well as $\lambda_0 = 5$ and $10$. As can be observed, $\chi$ does not affect the relation between $\lambda/\lambda_0$ and $n/n_0$ or $\lambda_M$. Also, $\lambda_0$ only affects the evolution of $n/n_0$ (DCM) at large contractions while it does not affect $\lambda_M^{-2}$ (CSM). From Eq. (20b) and (21b), we can deduce the scaling law

$$\frac{\lambda}{\lambda_0} \sim \left(\frac{n}{n_0}\lambda_M^{-2}\right)^{-1/3} \quad (\lambda, \lambda_0 \gg 1) \tag{22}$$

so that we have $\lambda/\lambda_0 \sim (n/n_0)^{-1/3}$, for DCM, and $\lambda/\lambda_0 \sim \lambda_M^{2/3}$, for CSM, making all log-log plots collapse on a line for $\lambda, \lambda_0 \gg 1$, *i.e.*, for small contraction. We can also observe that $n$ diverges to infinity as $\lambda$ approaches unity, since the denominator of the right-hand side of Eq. (20a) approaches zero. This occurs when $\lambda/\lambda_0$ approaches $1/\lambda_0 = 0.1$ (0.2) for $\lambda_0 = 10$ (5). The scaling law in Eq. (22) shows also that the effect of an increment in crosslink density, via DCM, and that of a shortening of the average chain length, via CSM, are qualitatively similar.

Let us now compute the free energy of the gel in chemo-mechanical equilibrium. Separating the elastic and the mixing energy contributions, from Eq. (3), (4), (6), and (22) we have

$$\psi_e = \frac{n}{2}[\lambda^2 \lambda_M^{-2} + 2\lambda_0^2 \lambda_M^{-2} - 2\ln(\lambda \lambda_0^2 \lambda_M^{-3}) - 3] \tag{23a}$$

$$\sim \frac{n_0}{2} \lambda_0^2 \left[\left(\frac{\lambda}{\lambda_0}\right)^{-1} + 2\left(\frac{\lambda}{\lambda_0}\right)^{-3}\right] \quad (\lambda, \lambda_0 \gg 1) \tag{23b}$$

and

$$\psi_m = (\lambda \lambda_0^2 - 1)\ln(1 - \lambda^{-1}\lambda_0^{-2}) - \chi \lambda^{-1}\lambda_0^{-2} \tag{24a}$$

$$\sim n_0 \lambda_0^2 \left(\frac{\lambda}{\lambda_0}\right)^{-1} - 1 \quad (\lambda_0 \gg 1) \tag{24b}$$

where $n$ and $n_0$ are taken from Eq. (20) and (21), respectively. Now the total free energy, from Eq. (1), can be obtained by summing the elastic and mixing contributions as $\psi = \psi_e + \psi_m$.

*Motor-limited regime*

In the motor-limited regime, *i.e.* at slow motor (SM) activation, we assume solvent flow is much faster than motor activation and, thus, occurs instantaneously, leaving the gel in chemo-mechanical equilibrium as the microstructure evolves. In this case, by neglecting the viscous dissipation created by solvent flow, the mechanical work $w \, kT/\Omega$ performed by the motors, per unit reference volume, is equivalent to the change in free energy in the gel, so that $w = \psi - \psi_0$. The power generation, per unit reference volume, is $\dot{w} = \partial w/\partial t$, so that

$$\frac{t}{t_0} = \frac{\psi - \psi_0}{n_0 \lambda_0^2} \tag{25a}$$

$$\sim \frac{3}{2}\left(\frac{\lambda}{\lambda_0}\right)^{-1} + \left(\frac{\lambda}{\lambda_0}\right)^{-3} - \frac{5}{2} \quad (\lambda, \lambda_0 \gg 1) \tag{25b}$$

with

$$t_0 = \frac{n_0 \lambda_0^2}{\dot{w}} \tag{25c}$$

the nominal contraction time. Figure 3 reports the plot of the dimensionless contraction time $t/t_0$ versus the stretch ratio $\lambda/\lambda_0$ in the SM regime for the case of DCM and CSM, for $\lambda_0 = 5$ and $10$, and for $\chi = -1.1$ and $0.4$. At the preliminary stage of contraction all curves collapse into one curve, which evolves into a line and then diverges to a horizontal asymptote for DCM.

The slope of the tendency line emerges from the second (leading) terms in Eq. (23b) and (25b). Substituting Eq. (22) into (25b) we have

$$\frac{t}{t_0} \sim \frac{3}{2}\left(\frac{n}{n_0}\lambda_M^{-2}\right)^{1/3} + \frac{n}{n_0}\lambda_M^{-2} - \frac{5}{2} \quad (\lambda, \lambda_0 \gg 1) \tag{26}$$

where the second term on the right-hand side is the leading one at significant contraction. The scaling in Eq. (26) compares again the effects of DCM and CSM, showing their qualitative similarity. In Figure 3 we can also observe that $\chi$ does not affect our results and that $\lambda_0$ is only effective at large contractions for DCM, as observed in Figure 2.

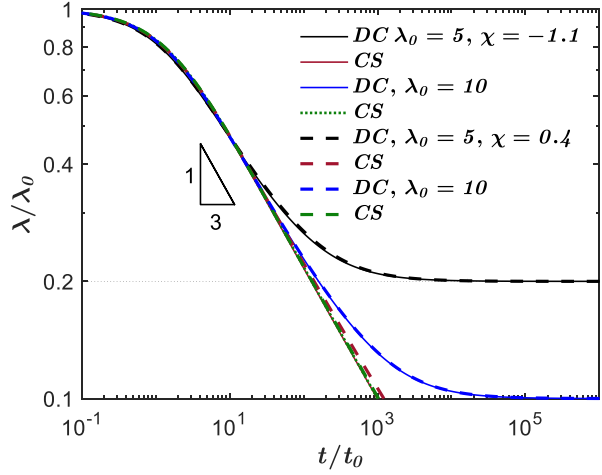

**Figure 3**: Motor-limited regime: Evolution of the contracted stretch ratio $\lambda/\lambda_0$ as a function of the dimensionless time $t/t_0$, with $t_0$ the nominal time reported in Eq. (25c), for the same model parameters as in Figure 2. We analyze the case of dynamic crosslinking (DC) motors as well as chain shortening (CS) motors.

Considering the gel parameters from [6], for which $\chi = -1.1$, $\lambda_0 = 10$, $n_0 = 1.61 \cdot 10^{-5}$, where the gel prototype contracted by 90% of its initial volume, thus $\lambda/\lambda_0 = 0.1$, within the timeframe of one hour, thus $t = 3600\ s$, the energetic demand can be calculated as follows. Considering slow motor activation (SM), we have $t/t_0 \sim (\lambda/\lambda_0)^{-3} = 10^3$. Now, since $t = 3.6 \cdot 10^3$, we have that $t_0 \sim 3.6\ s$. From Eq. (25c) we can deduce $\dot{w} = 4.47 \cdot 10^{-4}\ s^{-1}$. Taking $kT/\Omega = 7.1 \cdot 10^{28} kT\ m^{-3}$ [7], we can finally calculate the required power generation per unit reference volume, giving $\dot{w}\ kT/\Omega = 3.18 \cdot 10^7\ kT/\mu m^3 s$. The motor density is $N_M \sim 5 \cdot 10^5\ \mu m^{-3}$ [6-7], from which we obtain the required power generation per motor as $\sim 63.6\ kT/s$, which is equivalent to the hydrolysis of $\sim 6$ ATP molecules per second in each motor (one ATP molecule releases $\sim 12\ kT$ when hydrolyzing).

*Diffusion-limited regime*

In the diffusion-limited regime, the time scale of motor activation is much shorter than that of solvent diffusion. In this case, we have fast motor (FM) activation, and we assume that the microstructure evolves instantaneously, thereby bringing the gel out of equilibrium in the active initial configuration (*i*). The absence of equilibrium creates an excess of osmotic pressure, which motivates the solvent outflow. Solvent outflow finally accommodates contraction, which recovers the equilibrium of the gel in the final configuration (*f*). The latter is defined by Eq. (20) and Figure 2. In configuration (*i*), we have $\lambda = \lambda_0$, giving $J = \lambda_0^3$, and $n > n_0$ for DCM or $\lambda_M^{-2} > 1$ for CSM [7]. In the configuration (*f*), we have $\lambda_f < \lambda_0$, $J = \lambda \lambda_0^2$, and $n$ or $\lambda_M$ identical as in configuration (*i*).

As elaborated in Appendix A, the contraction kinetics follows the law

$$\frac{\lambda}{\lambda_0} \approx 1 - \sqrt{\frac{t}{t_1}} \tag{27a}$$

for $\lambda_f \leq \lambda \leq \lambda_0$, *i.e.* for $t \leq t_f$, with $t_f$ the final contraction time and

$$t_1 = \frac{L^2 \lambda_0^2}{D \beta^2} \tag{27b}$$

the characteristic time. Here, $D$ is the diffusivity of the solvent in the gel, and $\beta$ is a coefficient that depends on $n/n_0$ for DCM and $\lambda_M$ for CSM. This coefficient is calculated numerically and fitted to the simple law

$$\beta = B \gamma^b \tag{28}$$

where $\gamma = n/n_0$ for DCM, $\gamma = \lambda_M^{-2}$ for CSM, and $B$ and $b$ are fitting coefficients reported in Table 1, as function of $\lambda_0$ and $\chi$.

Figure 4 plots the contraction kinetics from the numerical results reported in Appendix A, and estimated by Eq. (30)

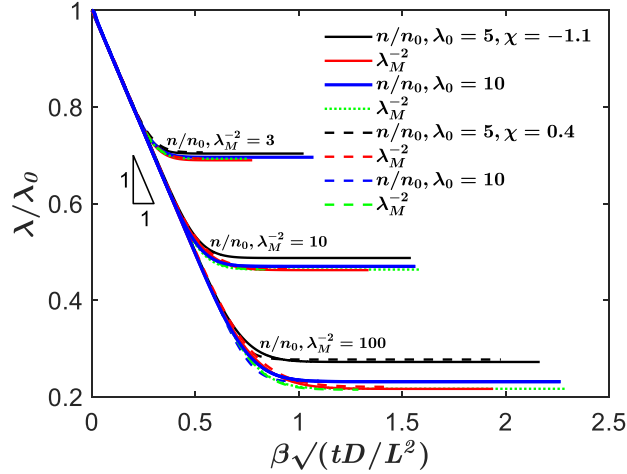

**Figure 4**: Diffusion-limited regime: Evolution of the contracted stretch ratio $\lambda/\lambda_0$ as a function of the dimensionless time $t/t_1$, with $t_1$ the nominal time reported in Eq. (27b), for the same model parameters as in Figure 2.

By inverting Eq. (30) we can estimate the final contraction time as

$$\frac{t_f}{t_1} \approx \left(1 - \frac{\lambda_f}{\lambda_0}\right)^2 \tag{29a}$$

$$\sim \left[1-\left(\frac{n}{n_0}\lambda_M^{-2}\right)^{-1/3}\right]^2 \quad (\lambda, \lambda_0 \gg 1) \tag{29b}$$

Consider now that the motor-limited regime occurs only for very low power generation by the motors, while the diffusion-limited regime occurs at very high power generations. To calculate the transition value for power generation we need to consider $t_{SM} \ll t_{FM}$, where $t_{SM}$ is taken from Eq. (28) or (26), while $t_{FM}$ is taken from Eq. (30) and (32). Considering the leading order terms, where $\lambda, \lambda_0 \gg 1$ (with $\lambda = \lambda_f$), we have

$$\frac{t_1}{t_0} \gg \left(\frac{\lambda}{\lambda_0}\right)^{-3} \sim \left(\frac{n}{n_0}\lambda_M^{-2}\right) \tag{30}$$

**Table 1**: Coefficients of Eq. (28)

|  | DCM | | CSM | |
| --- | --- | --- | --- | --- |
|  | $B$ | $b$ | $B$ | $b$ |
| $\lambda_0 = 5, \chi = -1.1$ | $7.25 \cdot 10^{-3}$ | 0.70 | $6.50 \cdot 10^{-3}$ | 0.74 |
| $\lambda_0 = 10$ | $8.2 \cdot 10^{-4}$ | 0.71 | $7.05 \cdot 10^{-4}$ | 0.74 |
| $\lambda_0 = 5, \chi = 0.4$ | $1.43 \cdot 10^{-3}$ | 0.72 | $1.44 \cdot 10^{-3}$ | 0.76 |
| $\lambda_0 = 10$ | $1.7 \cdot 10^{-4}$ | 0.73 | $1.75 \cdot 10^{-4}$ | 0.74 |

By substituting from Eq. (25c) and (27b) we obtain the transition power density

$$\dot{w}^* \approx \frac{D}{L^2} n_0 \beta^2 \left(\frac{\lambda}{\lambda_0}\right)^{-3} \tag{31}$$

so that $\dot{w} \gg \dot{w}^*$ provides the diffusion-limited regime, while $\dot{w} \ll \dot{w}^*$ gives the condition for the motor-limited regime. As observed in Eq. (31), the transition power density increases with diffusivity $D$ and reduces with slab reference thickness $L$. *I.e.*, for a given power density $\dot{w}$, a thinner gel slab or with higher solvent diffusivity (smaller average solvent molecules) is more likely to contract in the motor-limit regime (SM) since solvent diffusion will likely occur faster than motor activation.

**Conclusions**

The proposed model provides the contraction time for cytoskeletal gel slabs for the two extreme cases of slow motor activation (SM), or motor-limited contraction, and fast motor activation (FM), or diffusion-limited contraction. In the SM regime, the contraction time scales as $t/t_0 \sim (\lambda/\lambda_0)^{-3}$ (when $\lambda, \lambda_0 \gg 1$) with the nominal time $t_0$ proportional to $1/\dot{w}$, with $\dot{w}$ the power density of the motors per unit reference volume. In the FM regime, the contraction time scales as $t/t_1 \sim (1 - \lambda/\lambda_0)^2$, with the nominal time $t_1$ proportional to $L^2/D$, with $L$ the slab thickness in the reference (dry polymer) state, and $D$ the diffusivity of the solvent. Because the contraction time in the FM regime is always smaller than that in the SM regime, we obtain a transition power density $\dot{w}^*$, for which SM apply when $\dot{w} \ll \dot{w}^*$, while FM applies when $\dot{w} \ll \dot{w}^*$. Also, $\dot{w}^*$ is proportional to $D/L^2$, so that a thin gel slab with high diffusivity (smaller average solvent molecules) is more likely to contract in the motor-limited regime thanks to a faster solvent diffusion. For intermediate regimes where $\dot{w} \sim \dot{w}^*$, one has to solve the transient coupled problem with motor activation and solvent diffusion.

Our analysis provides important scaling arguments to define the contraction kinetics of active cytoskeletal gel slabs and can be used to observe molecular scale kinetic phenomena from macroscopic observations on gel contractions.

**Acknowledgments**

This work was supported by the New Frontiers in Research Funds – Exploration (NFRFE-2018-00730) and by the Natural Sciences and Engineering Research Council of Canada (NSERC) (RGPIN-2017-04464).

**Appendix A**

In the diffusion-limited regime, we solve the transient problem of solvent diffusion assuming the microstructure evolves instantaneously. The initial gel configuration is that of (*i*), where $\lambda = \lambda_0$, and $J = \lambda_0^3$, but where $n > n_0$ and $\lambda_M^{-2} = 1$, for DCM, or $n = n_0$ and $\lambda_M^{-2} > 1$ for CSM. Because the microstructural properties of $n$ and $\lambda_M^{-2}$ do not satisfy the chemo-mechanical equilibrium in Eq. (20), the chemical potential is non-uniform in the gel and will then generate a nonzero solvent flux $H$. Eq. (19) provides a partial differential equation for the function $\lambda(\tilde{t}, \tilde{Z})$, with $\tilde{t} = t/t_d$ and $\tilde{Z} = Z/L$ the dimensionless time and spatial coordinate, where $t_d = L^2/D$ is the characteristic diffusion time, with $L$ the reference thickness of the gel slab and $D$ the diffusivity of the solvent. The solution of Eq. (19) can be approximated to

$$\frac{\lambda}{\lambda_0} \approx 1 - \frac{\beta}{\lambda_0}\sqrt{\tilde{t}} \tag{A1}$$

for $\lambda < \lambda_f$, where $\lambda_f$ is the equilibrium stretch obtained from Eq. (20) with $n > n_0$ (for dynamic crosslinking motors, DCM) or $\lambda_M^{-2} > 1$ (chain shortening motors, CSM). In Eq. (A1), $\beta$ is a coefficient that depends on the microstructural features of the gel, *i.e.*, $n/n_0$ and $\lambda_M^{-2}$, as well as the rest state (0) of the gel given by $\lambda_0$, $n_0$ and $\chi$. Rearranging Eq. (A1) we can obtain Eq. (27), where $\beta$ is fitted to the function in Eq. (28) to the least square fit with a $R^2 \geq 0.99$. Eq. (28), *i.e.*, $\beta = B\,\gamma^b$, with $\gamma = n/n_0$ for DCM and $\gamma = \lambda_M^{-2}$ for CSM, has fitting coefficient $B$ and $b$ reported in Table 1. In Figure A1 we report the fitting of $\beta$ for DCM, $\lambda_0 = 10$, $\chi = -1.1$, and thus $n_0 = 1.61 \cdot 10^{-5}$.

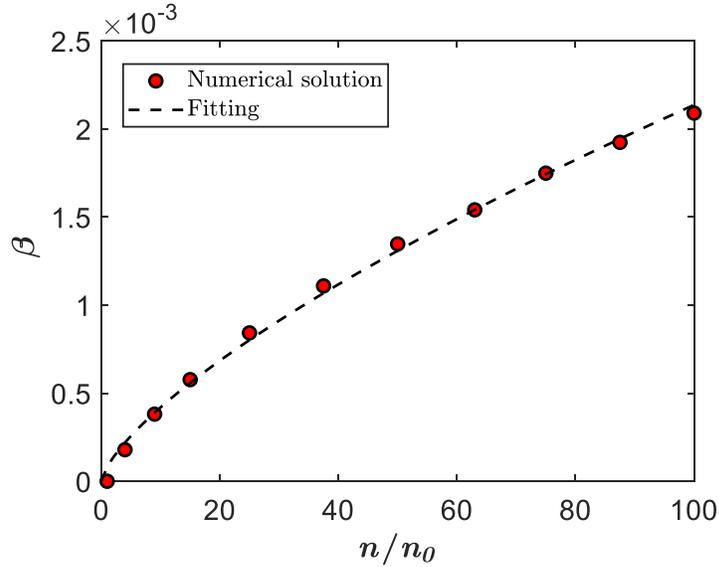

**Figure A1**: Numerical evaluation of the coefficient $\beta$ versus $n/n_0$ for dynamic crosslinking motors (DCM) with $\lambda_0 = 10$, $\chi = -1.1$, and thus $n_0 = 1.61 \cdot 10^{-5}$. The numerical solutions are fitted to the function $\beta = B(n/n_0)^b$ reported in Eq. (28), with the fitting coefficients $B$ and $b$ reported in Table 1.

## References

[1] Lieleg O, Ribbeck K (2011). Biological hydrogels as selective diffusion barriers. *Trends in Cell Biology*, 21(9):543–551.

[2] Boal DH (2002). Mechanics of the cell. *Cambridge University Press*.

[3] Avila J (1991). Minireview: microtubule functions. *Life Sciences*, 50 (1):327–334.

[4] Cooper GM (2000). The Cell: A molecular approach 2nd edition. *Sinauer Associates.*

[5] Gardel ML, Shin JH, MacKintosh FC, Mahadevan L, Matsudaira P, Weitz DA (2004). Elastic Behavior of Cross-Linked and Bundled Actin Networks. *Science*, 304(5675):1301–1305.

[6] Bendix PM, Koenderink GH, Cuvelier D, Dogic Z, Koeleman BN, Brieher WM, Field CM, Mahadevan L, Weitz DA (2008). A quantitative analysis of contractility in active cytoskeletal protein networks. *Biophysical Journal*, 94(8):3126–3136.

[7] Ferraresso M, Kong A, Hasan M, Agostinelli D, Elfring GJ (2022). Energetics of Cytoskeletal Gel Contraction. *Soft Matter, accepted.*

[8] Bacca M, Saleh OA, Mcmeeking RM (2019). Contraction of polymer gels created by the activity of molecular motors. *Soft Matter*, 22(15):4467-4475.

[9] Mackintosh FC, Levine AJ (2008). Nonequilibrium mechanics and dynamics of motor-activated gels. *Physical Review Letters*, 100(1):018104-1.

[10] Hatwalne Y, Ramaswamy S, Rao M, Simha RA (2004). Rheology of Active-Particle Suspensions. *Physical Review Letters*, 92 (11):118101.

[11] Deshpande VS, Mcmeeking RM, Evans AG (2006). A bio-chemo-mechanical model for cell contractility. *Proceedings of the National Academy of Sciences,* 103(38):14015-14020.

[12] Deshpande VS, McMeeking RM, Evans AG (2007). A model for the contractility of the cytoskeleton including the effects of stress-fibre formation and dissociation. *Proceedings of the Royal Society A: Mathematical, Physical and Engineering Sciences*, 463(2079):787–815.

[13] Liu T (2014). A constitutive model for cytoskeletal contractility of smooth muscle cells. *Proceedings of the Royal Society A*, 470:20130771.

[14] Hill AV (1938). The heat of shortening and the dynamic constants of muscle. *Of The Royal Society B,* 126(843):136-195.

[15] Huggins ML (1941). Solutions of long chain compounds. *Journal of Chemical Physics*, 9(5):440-440.

[16] Flory PJ (1942). Thermodynamics of high polymer solutions. *Journal of Chemical Physics*, 10(1):51–61.

[17] Bacca M, Mcmeeking RM (2017). A viscoelastic constitutive law for hydrogels. *Meccanica*, 52:3345–3355.

[18] Hong W, Zhao X, Zhou J, Suo Z (2008). A theory of coupled diffusion and large deformation in polymeric gels. *Journal of the Mechanics and Physics of Solids*, 56(5):1779–1793.

[19] Bertrand OJN, Fygenson DK, Saleh OA (2012). Active, motor-driven mechanics in a DNA gel. *Proceedings of the National Academy of Sciences of the United States of America*, 109(43):17342–17347.

[20] Koenderink GH, Dogic Z, Nakamura F, Bendix PM, MacKintosh FC, Hartwig JH, Stossel TP, Weitz DA (2009). An active biopolymer network controlled by molecular motors. Proceedings of the National Academy of Science, 106(36):15192-7

[21] Curatolo M, Nardinocchi P, Teresi L (2021). Mechanics of active gel spheres under bulk contraction. *International Journal of Mechanical Sciences*, 193:106147.

[22] Curatolo M, Gabriele S, Teresi L (2017). Swelling and growth: a constitutive framework for active solids. *Meccanica*, 52(14):3443–3456.

[23] Cervantes IC, Curatolo M, Nardinocchi P, Teresi L (2022). Morphing of soft structures driven by active swelling: a numerical study. *International Journal of Non-Linear Mechanics*. 141:103951